\documentclass[
   final            
  ]
{aipproc}
\layoutstyle{6x9}

\def\asec{\ifmmode ^{\prime\prime}\else$^{\prime\prime}$\fi}
\def\amin{\ifmmode ^{\prime}\else$^{\prime}$\fi}
\def\farcm{\hbox{$.\!\!^{\prime}$}}  

\begin{document}

\title{Discovery of the Faint Near-IR Afterglow of GRB
  030528\footnote{for the GRACE collaboration}}

\author{Rau A.}{
  address={Max-Planck Institute for extraterrestrial Physics, Garching
 , Germany}
}

\author{Greiner J.}{
  address={Max-Planck Institute for extraterrestrial Physics, Garching
 , Germany}
}

\author{Klose S.}{
  address={Th\"uringer Landessternwarte, Tautenburg, Germany}
}

\author{Castro Cer\'on J. M.}{
  address={Space Telescope Science Institute, Baltimore, USA}
}

\author{Fruchter A.}{
  address={Space Telescope Science Institute, Baltimore, USA}
}

\author{K\"{u}pc\"{u} Yolda\c{s} A.}{
  address={Max-Planck Institute for extraterrestrial Physics, Garching
 , Germany}
}

\author{Gorosabel J.}{
  address={Space Telescope Science Institute, Baltimore, USA}
}

\author{Levan A. J.}{
  address={Space Telescope Science Institute, Baltimore, USA}
}

\author{Rhoads J. E.}{
  address={Space Telescope Science Institute, Baltimore, USA}
}

\author{Tanvir N. R.}{
  address={Department of Physical Science, Univ. of Hertfordshire,
  Hatfield Herts, UK}
}


\begin{abstract}
We report on the discovery of the near-IR transient of the
long-duration gamma-ray burst GRB 030528 and its underlying host
galaxy. The near-IR transient was first observed in $JHKs$ with SofI
at the 3.6\,m ESO-NTT 16\,hrs after the burst and later observations
revealed a fading $Ks$-band afterglow. The afterglow nature was
confirmed by {\it Chandra} observations which found the source to be a
fading X-ray emitter. The lack of an optical afterglow and the early
faintness in the near-IR ($Ks$$>$18.5 mag) place GRB 030528 in a
parameter space usually populated by dark bursts. We find the host to
be an elongated blue galaxy.
\end{abstract}

\maketitle


\section{Observations}

On May 28 2003 {\it HETE-2} localized a moderately bright
(peak flux of 4.8$\times$10$^{-8}$ erg/cm$^2$/s at 30-400\,keV)
long-duration Gamma-ray burst (GRB) \cite{Atteia:2003} 107\,min after
the observed prompt emission (13:03 UT). Shortly after, ToO observations of the
circulated 2\amin\ radius error circle were initiated with the the
3.6\,m ESO-New Technology Telescope (NTT) equipped with the Son of
ISAAC (SofI) infrared spectrograph and imaging camera at La
Silla/Chile. Here we report on $JHKs$ imaging data of the first three nights
supplemented by later $I$-, $Js$- and $K$-band imaging with the
Mosaic2 imager at the 4\,m Blanco Telescope at the Cerro Tololo
Inter-American Observatory (CTIO), the Infrared Spectrometer And Array
Camera (ISAAC) at the 8.2\,m Very Large Telescope (VLT) in Paranal and
the 3.8\,m United Kingdom Infra-Red Telescope Fast-Track Imager (UKIRT
UFTI) on Mauna Kea, respectively (see Tab.1 for an observing log).

NTT-SofI observations were accomplished 16, 41 and 87\,hrs after
the GRB. Unfortunately, the {\it HETE-2} localization was revised to a
2\farcm5 radius error circle displaced by 1\farcm.3 from the
earlier position \cite{Villasenor:2003} after these images were
taken. Given the 5\farcm5 field of view of NTT-SofI, our data were
not covering the entire revised error circle. This, and the
non-detection of a transient in optical ($R$$>$18.7\,mag 2.3\,hrs
after the burst \cite{Ayani:2003}) and radio\cite{Frail:2003}
observations of the crowded field would have made the discovery of
the counterpart nearly impossible without {\it Chandra}
observations. Following the detection of four X-ray sources inside the
revised error circle on June 3 \cite{ButlerA:2003}
one of these sources (CXOU J170400.3--223710) was found to be fading in
our SofI-$Ks$-band images \cite{Greiner:2003}. A second {\it Chandra}
observation on June 9 showed a significant fading of the counterpart
candidate while the other sources inside the error circle
did not reveal any brightness decline \cite{ButlerB:2003}.

\begin{table}[t]
\begin{tabular}{crcccc}
\hline
\tablehead{1}{c}{b}{Instrument}
  & \tablehead{1}{r}{b}{Date (Start UT)}
  & \tablehead{1}{c}{b}{Filter}
  & \tablehead{1}{c}{b}{Exp. [min]}
  & \tablehead{1}{c}{b}{Seeing}
  & \tablehead{1}{c}{b}{Magnitude} \\
\hline
NTT-SofI & 29.5. 04:58 & J & 15 & 0.8\asec & 20.9$\pm$0.9 \\
NTT-SofI & 29.5. 05:16 & H & 15 & 0.8\asec & 19.9$\pm$0.9 \\
NTT-SofI & 29.5. 05:32 & Ks & 15 & 0.8\asec & 18.5$\pm$0.4 \\
NTT-SofI & 30.5. 04:54 & J & 20 & 1.6\asec & $>$20.1$\pm$0.7 \\
NTT-SofI & 30.5. 05:16 & H & 20 & 1.1\asec & $>$18.9$\pm$1.6 \\
NTT-SofI & 30.5. 05:40 & Ks & 20 & 1.1\asec & 18.8$\pm$0.3 \\
NTT-SofI & 1.6. 04:07 & Ks & 60 & 0.8\asec & 19.5$\pm$0.6 \\
Blanco-Mosaic2 & 4.6. 02:10 & I & 40 & 1.2\asec & 21.4$\pm$0.5 \\
UKIRT-UFTI & 12.6. 08:55 & K & 116 & 0.6\asec & 19.3$\pm$0.2 \\
Blanco-Mosaic2 & 30.6. 03:00 & I & 40 & 1.1\asec & 21.4$\pm$0.5 \\
VLT-ISAAC & 17.9. 00:12 & Js & 50 & 0.6\asec & 21.3$\pm$0.4 \\
VLT-ISAAC & 27.9. 00:27 & Js & 60 & 0.9\asec & 21.4$\pm$0.4 \\
VLT-ISAAC & 29.9. 23:38 & Js & 94 & 0.7\asec & 20.9$\pm$0.4 \\
VLT-ISAAC & 1.10. 00:04 & Js & 60 & 0.6\asec & 21.0$\pm$0.6 \\
\hline
\end{tabular}
\caption{Observation log. 
}
\label{tab:log}
\end{table}


The NTT-SofI and VLT-ISAAC data were reduced using the {\it Eclipse}
package\cite{Devillard:1997}. The reduction of the Blanco-Mosaic2 data
was performed with {\it bbpipe}, a script based on the {\it
IRAF/MSCRED}.
The photometry was performed with {\it IRAF/DAOPHOT}. We
calibrated the $JJsHK$ and $Ks$ images against several stars contained
in the 2MASS All-Sky Point Source
Catalog\footnote{http://irsa.ipac.caltech.edu/applications/Gator/}
from the neighbourhood of the GRB. The Mosaic2 I-band images were
calibrated using USNOFS field photometry\cite{Henden:2003}.
The magnitudes are corrected for foreground
extinction\cite{Schlegel:1998} (A$_K$=0.22, A$_H$=0.35, A$_J$=0.54 and
A$_I$=1.17) for the given galactic coordinates.

\section{The Near-IR Afterglow}

Observations from the first night ($\sim$16\,hrs after the burst)
showed the afterglow to be near the detection limit of the $JH$-band
SofI images. The source is well visible in $Ks$ at an extinction
corrected magnitude of 18.5$\pm$0.4. It was thus $\sim$1.5\,mag
brighter than the faintest $K$-band\footnote{We assume here
$K$$\sim$$Ks$, which is justified within the error of the estimated
magnitudes.} afterglow observed at that time after the burst, that of
GRB 971214.

\begin{figure}
  \includegraphics[height=.53\textheight,angle=-90]{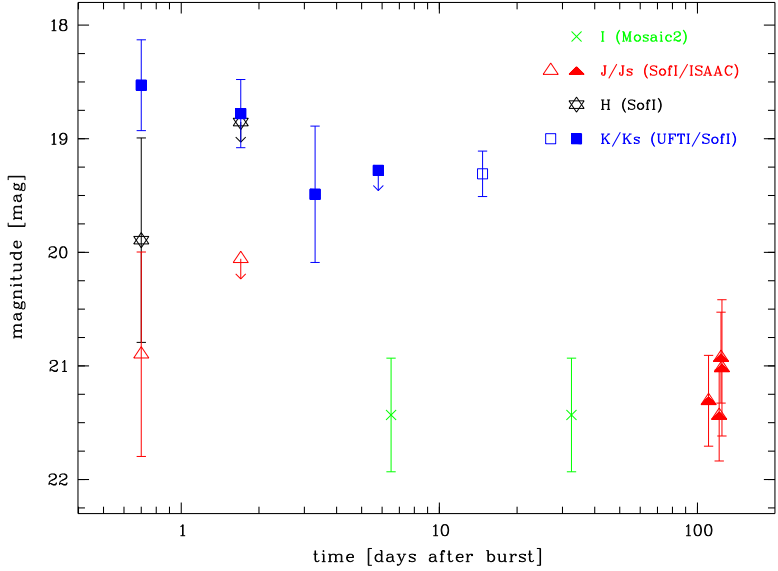}
  \caption{Light curves in $I$ (marked by a cross), $J$/$Js$
  (triangle/filled triangle), $H$ (star) and $K$/$Ks$ (square/filled
  square). For $J$ and $H$ only upper limits exist for
  t=1.7\,days. The $Ks$-band upper limit at t=5.8\,days is from
  \cite{bogosavljevic:2003}. All magnitudes are corrected for
  foreground extinction in the Galaxy.}
  \label{fig:lightcurves}
\end{figure}

The afterglow shows a probable fading in $Ks$ from 16 to 80\,hrs after
the burst by $\sim$1\,mag (Fig.~\ref{fig:lightcurves}). This
corresponds to a decay with a slope of
$\alpha$$\sim$0.6
$Ks$/$K$-band data from June 1 (NTT) and June 12 (UKIRT) shows
constant magnitudes. Therefore, the host galaxy seems to dominate the
$K$-band emission already after $\sim$3\,days. Unfortunately, the
observing conditions from the second night (41\,hrs after the burst) make
the estimation of an early fading in $J$ and $H$ not feasible. No
decay is seen in the $J$/$Js$-band\footnote{The $J$ and $Js$ filters
differ slightly in width and efficiency. For a flat spectrum a first
order assumption of $J$$\sim$$Js$+0.4\,mag can be used.} when
comparing the SofI observations and the ISAAC observations from
$>$100\,days. This suggests, that the afterglow emission was strongly
extinguished in the $J$-band (in addition to the galactic extinction)
and the emission dominated by the host galaxy already at the time of
our first SofI observation.

It is unlikely that we would have found this IR counterpart without
{\it Chandra}, and the non-detection in the optical would have led to
the classification of GRB as a dark burst. While $\sim$60\% of all
well localized GRBs have no detected optical/NIR afterglow, the fast
reaction time and deep detection limit of $\sim$19.5 in $Ks$  were
important for the discovery of the near-IR transient of GRB 030528.

\section{The Host Galaxy}
\begin{figure}
  \includegraphics[height=.33\textheight]{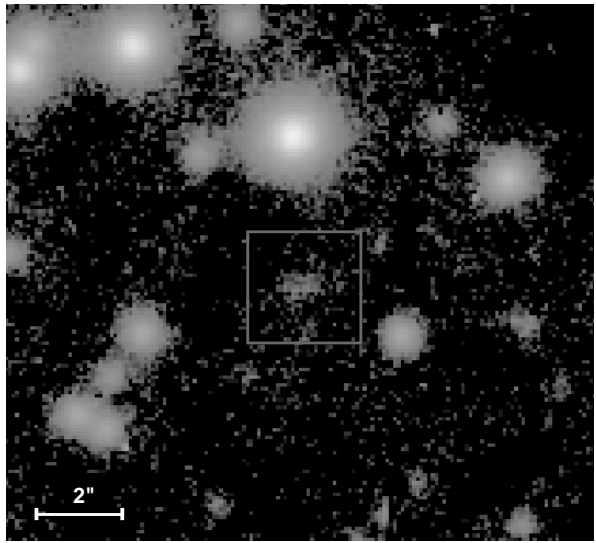}
  \caption{$K$-band image taken with the 3.8m UKIRT equipped with the
  UFTI on June 12, 15\,days after the prompt emission. North is up and
  East to the right. The potential host galaxy of GRB 030528 (square)
  shows an elongation in East-West direction.}
  \label{fig:Khost}
\end{figure}

Late time $K$-band imaging with UKIRT revealed the potential
underlying host galaxy (see Fig.~\ref{fig:Khost}). The data show an
East-West elongated object at the position of the transient with a
size of $\sim$1.5\asec$\times$0.8\asec. This shape suggests the host
to be either an elliptical or edge-on disk galaxy.
The apparent brightness of $K$$\sim$19.5\,mag and $I$-$K$$\sim$2\,mag
places the host of GRB 030528 well in the mean of the sample of GRB
host galaxies detected in the $K$-band \cite{leFloch:2003} and is more
typical for a star-forming rather than for an elliptical galaxy.
 
\begin{theacknowledgments}
This work is primarily based on observations collected at the European
Southern Observatory, Chile, under the proposal 71.D-0355 (PI:
E.v.d.Heuvel) with aditional data obtained at the Cerro Tololo
Inter-American Observatory and the United Kingdom Infra-Red
Telescope. This publication makes use of data products from the Two
Micron All Sky Survey, which is a joint project of the University of
Massachusetts and the Infrared Processing and Analysis
Center/California Institute of Technology, funded by the National
Aeronautics and Space Administration and the National Science
Foundation.
\end{theacknowledgments}





\end{document}